\documentstyle[aps,prc,preprint,tighten,dina4]{revtex}

\begin{document}
\title{Two-Photon Decays of Mesons in a Relativistic Quark Model}
\author{Claus\,R.\,M\"unz\footnote{E-mail: muenz@itkp.uni-bonn.de}}
\address{Institut f\"ur Theoretische Kernphysik,\\
         Universit\"at Bonn, Nu{\ss}allee 14-16, D-53115 Bonn, Germany}
\date{\today}
\preprint{\vbox{\hspace{-0.6cm}Bonn TK-96-01 \hfill Submitted to
    Physical Review D}} 
\maketitle

\begin{abstract}
  We present a relativistic calculation of two-photon decays for heavy and light
  mesons in the framework of the Salpeter equation for quark-antiquark states.
  The meson-photon-photon vertex is computed by reconstructing the
  Bethe-Salpeter vertex function and evaluating the four-dimensional Feynman
  diagram with off-shell quark amplitudes.

  The two-photon width for light and heavy quarkonia up to spin equal six are
  calculated with different parameter sets taken from the literature thus giving
  a complete overview on the mesonic two-photon physics.  We find that
  relativistic effects including the negative energy components of the wave
  function are important for any two-photon width - even for heavy quarkonia -
  yielding a remarkable agreement with available data.

\end{abstract} \pacs{}

\narrowtext

\section{Introduction} \label{I}

The decay into two photons is considered as an interesting experimental
playground in the mesonic physics of the near future.  New data
will become available not only from existing experiments at CLEO, LEP, TRISTAN
and VEPP, see~\cite{MPW} for a recent review on present experimental data, but
from upgrades such as LEP2000 and new facilities as DA$\Phi$NE or the Cornell,
KEK and SLAC B-factories.

The two-photon decay of mesons can be used to identify the flavor of
quark-anti\-quark states, but it may also discriminate between conventional
mesons and glueballs or hybrids. Especially in the isoscalar spectrum of light
scalar and tensor mesons a variety of resonances has been found which does not
fit into the conventional $q\bar q$ nonets. A quantitative theoretical and
experimental understanding of the two-photon decays may help to interpret the
meson spectrum.

Most of the few theoretical results, however, contain large uncertainties in
the overall scale of two-photon widths, especially in the light quark sector.
The reason is, that the corrections to nonrelativistic decay formulae are large
even in case of heavy quarkonia~\cite{Ber79,Ber83,Li91}. The situation is most
dramatic in the case of the pion, where a nonrelativistic ansatz fails by
orders of magnitude. Relativistic corrections up to now are based on the
Feynman diagram for a {\it free} quark-antiquark pair annihilating into two
photons.  Bergstr\"om et al.~\cite{Ber83} keep only positive energy components
of the intermediate quark propagator and use nonrelativistic wave functions.
Their results especially in the light meson sector do not agree well with
experimental data. Li et al.~\cite{Li91} use a field-theoretical approach
from the decay of p-wave positronium. From simple harmonic oscillator wave
functions they calculate relativistic corrections to the nonrelativistic ratio
15/4 between the scalar and tensor decay width. The authors stress the
necessity for a complete study within quark models, which give a good
description of meson spectroscopy.

The most complete references on the calculation of two-photon widths are those
of Godfrey and Isgur~\cite{GI} and of Ackleh and Barnes~\cite{AB92}. They use
wave functions, which follow from a semi- or nonrelativistic calculation of the
meson mass spectra, which again is multiplied with the Feynman transition
amplitude for a {\it free} quark-anti\-quark pair annihilating into two
photons. As it stands, this would, however, not at all give the correct
dependence of the widths on the meson bound-state mass. Therefore for instance
in the case of pseudoscalar mesons an additional phenomenological factor
$(M_{{exp}}/M_{{ref}})^3$ is introduced, where $M_{{exp}}$ is the experimental
(or calculated) bound-state mass and $M_{{ref}}$ is a reference mass, which
in~\cite{GI,AB92} is taken to be the rest plus kinetic energy of the
quark-anti\-quark pair in the meson (``Mock-meson mass''). This
phenomenological input is necessary, because the transition is calculated
between free quarks, which especially in the case of a deeply bound such as the
pion is a very crude approximation. However, it seems that such a mass
dependence is a crucial input, as most of their results agree well with
existing data.  As pointed out by the authors, there is considerable
arbitrariness and lack of a stringent derivation in the choice of the reference
mass, and hence a corresponding uncertainty in the overall scale of the decay
rates.

With this background we present a calculation of two-photon widths in the
framework of the instantaneous Bethe-Salpeter equation, which has several
improvements as compared to the existing theoretical estimates. On the one hand
we use the Salpeter equation for the calculation of the meson mass spectrum,
which includes the negative energy components of the wave function.  This
allows for a relativistic normalization of quark-anti\-quark amplitudes given
by Salpeter~\cite{Sa}, which has a reasonable limit for deeply bound states.
On the other hand we calculate the decay matrix element in the framework of the
Mandelstam formalism, so that the dependence of the quark-anti\-quark relative
{\it four}-momentum is accounted for. In that way the transition amplitude from
the quark-antiquark state into two photons can be formulated for {\it
  off-shell} quarks. Taken together this leads to the correct mass
dependence of the decay widths and thus eliminates additional phenomenological
input of previous calculations.

The formalism is applied to a complete calculation of light and heavy mesons up
to spin equal six, for both singlet and triplet states. The comparison of
various model parameters given in the literature will lead to results, which
are relatively model independent or at least give a theoretical error, and
therefore may be used as a basis for experimental investigations and their
interpretation.

\section{Relativistic Calculation of Two-Photon Width in the Framework of the
  Salpeter Equation} \label{II}
\subsection{Spectra and Two-photon Widths in the Salpeter Formalism}
Starting point in our calculation of the two-photon decays is the
Bethe-Salpeter equation in its instantaneous approximation (Salpeter equation).
Its advantage with respect to nonrelativistic models are the inclusion of the
full Dirac-structure and negative energy components in the wave function, which
should be important for light quarkonia. Moreover, the instantaneous
interaction allows to incorporate a phenomenological confinement in the form of
a linearly rising potential. The equation for the Bethe-Salpeter amplitude
$\chi(p)$ in the rest frame of the meson reads
\begin{eqnarray}
\label{4}
S^{-1}(M/2+p)\,\chi^{}(p)\, S^{-1}(-M/2+p) &=&  \int\! \frac{d^4
  p'}{(2\pi)^4}\, [-i\,V(\vec p, \vec p \,')\,\chi^{}(p')]
\end{eqnarray} 
where we have used an instantaneous interaction $V(\vec p, \vec p \,')$ and the
quark propagators $S(p_i)$. Defining the equal time amplitude 
$\Phi^{}(\vec p\,):=\int\! {dp_0}/{2\pi}\,\chi^{}(p_0,\vec p\,)$, we arrive at
the well known Salpeter equation
\begin{eqnarray}
M\,\Phi(\vec{p}\,) & = & \;\; H(\vec{p}\,)\,\Phi(\vec{p}\,) -
  \Phi(\vec{p}\,)\,H(-\vec{p}\,) 
 \label{9}
\\
 &-&
\Lambda^+(\vec{p}\,)\,\gamma^0\;\int \!\!\frac{d^3p'}{(2\pi)^3}\,
[V(\vec{p},\vec{p}\,')\,\Phi(\vec{p}\,')]
\,\gamma^0\,\Lambda^-(-\vec{p}\,)
 \nonumber \\
 &+&
\Lambda^-(\vec{p}\,)\,\gamma^0\;
\int \!\!\frac{d^3p'}{(2\pi)^3}\,
[V(\vec{p},\vec{p}\,')\,\Phi(\vec{p}\,')]
\,\gamma^0\,\Lambda^+(-\vec{p}\,) 
 \nonumber 
\end{eqnarray} 
with the projectors \(\Lambda^{\pm} = (\omega \pm
H)/(2\omega)\), the Dirac Hamiltonian
\mbox{\(H(\vec{p}\,)=\gamma^0(\vec{\gamma}\vec{p}+m)\)} and
$\omega=(m^2+\vec{p\,}^2)^{1/2}$ with $m$ the quark mass.
The Salpeter equation is an eigenvalue equation for the bound-state mass $M$
and can be solved numerically, see for
instance~\cite{RMMP94,Mue94,RM95}. 
It is important to note, that the normalization of the Salpeter
amplitudes~\cite{Sa} 
\begin{eqnarray} 
       \int\! \frac{d^3p}{(2\pi )^3}
         \;\mbox{tr} \; \Big\{ \Phi^{\dagger}(\vec{p}\,) \Lambda^+(\vec{p}\,)
                          \Phi(\vec{p}\,)   \Lambda^-(-\vec{p}\,) 
   -         \Phi^{\dagger}(\vec{p}\,) \Lambda^-(\vec{p}\,)
                          \Phi(\vec{p}\,)   \Lambda^+(-\vec{p}\,)  
                    \Big\} =    \, 2M
\label{n1} 
\end{eqnarray}
is mandatory to obtain a reasonable dependence on the bound-state mass $M$.

The Bethe-Salpeter amplitude $\chi(p)$ is needed to calculate current matrix
elements of the corresponding bound-states in the Mandelstam
formalism~\cite{Ma}. As has been pointed out in~\cite{RMMP94},
equation (\ref{4}) allows for the reconstruction of the Bethe-Salpeter
amplitude $\chi(p)$ in the rest frame from the equal
time amplitude $\Phi(\vec p\,)$ as 
\begin{eqnarray} 
\chi^{}(p)  &=& S(M/2+p)\; \int\! \frac{d^3 p'}{(2\pi)^3}\,
[-i\,V(\vec p, \vec p \,')\,\Phi^{}(\vec p \,')]\; S(-M/2+p)
\end{eqnarray}

The transition amplitude for a meson with mass $M$ decaying
into two photons with momenta $k$ and $\widetilde{k} $ and polarization vectors 
$\varepsilon_1,\varepsilon_2$ then follows from the Mandelstam formalism as
\begin{eqnarray}
T & = & -i \sqrt{3}\; (ie_q)^2 \!
 \int \!\! \frac{d^4p}{(2\pi)^4}
              \; \mbox{tr}\; \Bigg\{ \,
    \chi(p) \left(\varepsilon\!\!\! /_2\, S(\frac{M}{2}+p-k)\,
    \varepsilon\!\!\! /_1 +  
     \varepsilon\!\!\! /_1\, S(\frac{M}{2}+p-\widetilde{k})\, 
    \varepsilon\!\!\! /_2 \right)
        \Bigg\} 
\label{tmat}
\end{eqnarray}
so that we find for the decay width
\begin{eqnarray}
  \lefteqn{    \Gamma(M\rightarrow \gamma\gamma)  = 
     \frac{3\pi}{2} \,\frac{\alpha^2}{M}\, \frac{1}{2J+1}}\\
  &  &
      \sum_{M_J, \lambda_1,\lambda_2}    
 \left|    
    \int \!\! \frac{d^4p}{(2\pi)^4} \,\frac{e_q^4}{e^4}\,
    \mbox{tr}\; \Bigg\{ \;
    \chi^{}_{M_J}(p) \left(\varepsilon\!\!\! /_2\; 
    S(\frac{M}{2}+p-k)\; \varepsilon\!\!\! /_1 + 
     \varepsilon\!\!\! /_1\; 
    S(\frac{M}{2}+p-\widetilde{k})\; 
    \varepsilon\!\!\! /_2 \right)
        \Bigg\}    \right| ^2
\nonumber
\end{eqnarray}
where $\lambda_1,\lambda_2$ are the polarizations of the two photons, $ J$ and
$M_J$ are the spin quantum numbers of the decaying meson and $k$ is
fixed for example to the positive z-direction.

As for the numerical treatment, we  have expanded the radial basis functions in
a set of twenty Laguerre functions in momentum space. The results for both
masses and two-photon widths are found to be stable with respect to a
variation of  the number of basis states and to the scale of the basis
functions.  

\subsection{Quark-Antiquark Interaction}
The most successful ansatz in parameterizing quark confinement in hadron
spectroscopy has been a linear potential, see for instance Godfrey and
Isgur~\cite{GI} for a detailed study of mesons as quark-antiquark
states. Motivated from lattice calculations for static quarks, its spin
structure is usually taken to be scalar. This however, must not be true for
light quarks. We therefore use a confinement spin structure
\begin{equation}
  \left[V_C(\vec{p},\vec{p}\,')\,\Phi(\vec{p}\,')\right] = {\cal
    V}_C(\vec{p}-\vec{p}\,')\;
  \left[(1-x)\,\Phi(\vec{p}\,')-x\,\gamma^0\,\Phi(\vec{p}\,')\,\gamma^0\right]
\label{conf}
\end{equation}
where $x=0$ is the scalar confinement used in non- and semi-relativistic quark
models and $x=1$ is a timelike vector spin structure used in the Salpeter model V
of Resag and M\"unz~\cite{RM95} for heavy quarkonia.  The scalar function
\({\cal V}_C\) is given in coordinate space by the commonly used linearly
rising potential \({\cal V}_C(r) = a+b r\).

For heavy quarkonia one usually takes beside the linear confinement potential a
residual interaction coming from the one-gluon-exchange (OGE). Its
implementation with a running coupling constant in the Salpeter formalism has
been discussed extensively in~\cite{RM95}.  In order to obtain results which
are as far as possible model independent, we compare in the following several
parameter sets for heavy quarkonia found in the literature. We investigate the
parameter set of a nonrelativistic calculation for both mesons and baryon of
Bhadhuri et al.~\cite{BC} with a fixed coupling constant and the
semi-relativistic ansatz of Godfrey and Isgur for the meson spectrum and
decays~\cite{GI} with running coupling constant. A fixed coupling constant
$\alpha_s$, however, leads in part to divergent Salpeter amplitudes for
$r\rightarrow 0$~\cite{Mur}. For a running coupling constant this phenomenon is
less pronounced, but still present. We therefore use a regularization procedure
by cutting off the OGE potential smoothly for $r < r_0$~\cite{RM95}. In the
model of~\cite{BC} we use $r_0 = 0.1\,$fm, in ~\cite{GI} we use their parameter
from the smearing function, which essentially is an equivalent regularization
procedure.  The meson masses calculated in the Salpeter model differ
substantially only for the $\eta_c$ in~\cite{BC}, which lies 100 MeV lower than
in the nonrelativistic calculation.  In addition to the parameters of this non-
and semi-relativistic quark model we compare the Salpeter model \cite{RM95} with
timelike vector confinement structure for heavy quarkonia.

For light quarkonia with u,\ d and s quarks the situation is different with the
following respects. We can no longer use parameter sets of non- or
semi-relativistic quark models, as the dynamics including the negative energy
components plays a quantitative effect. Moreover, the Salpeter equation with a
purely scalar confinement leads to an instability~\cite{Arch,Piek}, which
becomes prominent in the case of light mesons. We therefore use a
confinement structure with $x= 0.5$ to find reliable solutions for light
quarkonia and to describe the spin-orbit splitting of the triplet p-wave mesons
as good as possible.

In addition, the Salpeter equation does not allow to reproduce the masses of
the light and heavy quarkonia with the same confinement strength, as can be
done in a semi-relativistic ansatz~\cite{GI}. The radial excitations of the
$\Psi$ and $\Upsilon$ states require a confinement strength of typically
$b^{c,b}\approx 1300\,$MeV/fm~\cite{RM95,PJP95}, whereas the description of the
Regge trajectories for isovector and strange mesons requires $b^{u,d,s}\approx
1900\,$MeV/fm~\cite{PJP95,KMMP95}. Unfortunately the Salpeter models for light
mesons with OGE interaction that have been presented so far use a fixed
coupling constant $\alpha_s$~\cite{PJP95} with the above mentioned divergent
Salpeter amplitudes for $r\rightarrow 0$, which are not suitable for our
purpose. We thus present two fits for light quarkonia, which both
reproduce the experimental mass spectrum except for the $\eta$ meson sector. We
use a semi-relativistic value for the nonstrange quark mass of 220 MeV in the
fit OGE-SRM ({\em S}emi-{\em R}elativistic quark {\em M}ass) and a nonstrange
mass of 330 MeV common to nonrelativistic calculations in OGE-NRM ({\em
  N}on-{\em R}elativistic quark {\em M}ass). As shown in~\cite{AB92}, the quark
mass dependence of the two-photon widths is the most prominent one, so that the
two models will estimate the theoretical error.  As it is not possible to
formulate the one-gluon-exchange interaction gauge invariantly, we will use the
Feynman gauge in the first and the Coulomb gauge in the latter parameter set.
Again we use the regularization parameter of $r_0=0.1\,$fm.

In the light quark sector, however, 't Hooft~\cite{Hoo76} has derived another
QCD inspired residual $q\bar q$ interaction coming from instanton effects. It
has been proposed to solve the U$_A$(1) problem and was successfully used in a
nonrelativistic setup for the description of the $\pi$-$\eta$ splitting and the
$\eta$-$\eta'$ mixing~\cite{Bla90}. In the framework of the Salpeter equation
it acts in addition on scalar mesons and gives an interesting interpretation of
the scalar mixing and splitting~\cite{KMMP95}. We will therefore also present
two-photon widths in this model for comparison.  The parameter sets and
interaction type of all the above mentioned models are given in Table
\ref{tabparams}.

\section{Results and discussion} \label{III}
\subsection{Two-photon Decays of Heavy Quarkonia}
In Tables~\ref{tabggcc} and~\ref{tabggbb} we present our numerical results for
the two-photon widths of the charmonium and bottomonium singlet and triplet
states up to spin equal four which are allowed by the
Yang-theorem~\cite{Ya50}. We used the three above mentioned parameter sets 
to extract a theoretical estimate including a ``statistical'' error, which of
course includes only the parameter dependence. As the width is decreasing very
rapidly with increasing total spin for these (almost) nonrelativistic systems ,
higher angular excitations are no more interesting. These calculations complete
and improve results from other authors on spin singlet states~\cite{AB92} and
on scalar and tensor mesons~\cite{Ber83}. The conceptional improvement with
respect to previous work comes from the use of relativistic
amplitudes, from their Salpeter normalization and from the off-shell treatment
of the quarks, which allows to renounce on phenomenological ``Mock-meson''
correction factors, being more and more uncertain for angular excited states.
To estimate the effect of these improvements, we have compared in table
\ref{tabberg} our widths of the triplet charmonium and bottomonium states with
those of the work of Bergstr\"om et al.~\cite{Ber83} using the same parameter
set of \cite{BC}. For the scalar mesons their full result, which is a factor of
two smaller than the nonrelativistic ansatz, agrees with ours almost
quantitatively. The tensor states, however, seem to be more sensitive to the
above discussed relativistic effects, as our results differ from those
of~\cite{Ber83} by a factor of two.

Our calculation shows good agreement with available experimental
data~\cite{PDG94} on the $\eta_c$, $\chi_0$ and $\chi_2$, especially with a
recent measurement from the CLEO collaboration~\cite{Cleo95}. The results agree
well with those of the theoretical prediction of Ackleh and Barnes~\cite{AB92}
for the singlet bottomonium states. Their widths of singlet charmonium,
however, are somewhat larger.

As concerned to future investigations for instance in $\gamma\gamma$ collision
experiments our calculations suggest a strong coupling of the ground state and
radial excitations of the $\eta_c$ and $\chi_0$ charmonium states. Especially
the $\eta_c'(3590)$ with a predicted width of $\Gamma(\eta_c'\rightarrow
\gamma\gamma)=1400\pm 400$ eV could probably be confirmed in a $\gamma\gamma$
experimental setup, as it lies below the $D\bar D$ threshold.  In the
bottomonium sector only the $\eta_b$ has an appreciable two-photon width of 200
eV beside its radial excitations lying below the $B\bar B$ threshold with
a width of around 100 eV.

\subsection{Two-photon Decays of Light Quarkonia}
In the light quark sector a relativistic treatment of meson decay observables
becomes mandatory, see for instance~\cite{Mue94b} for an almost quantitative
description of the electromagnetic decays and form factors of the light ground
state mesons.

One of the major conceptional improvements as compared to previous models is
the dependence of the two-photon width on the bound-state mass $M$. If one uses
the simplest possible Lagrangian for the coupling of a pseudoscalar field
$\phi$ to two photons
\begin{equation}
 {\cal L} = \frac{1}{2} \, g\, \phi F_{\mu\nu}\, \widetilde F^{\mu\nu}
\label{lagr}
\end{equation}
one finds for the width \cite{AB92}
\begin{equation}
 \Gamma(M\rightarrow\gamma\gamma) = \frac{1}{64\pi}\, g^2\,M^3.
\label{width}
\end{equation}
To check this behavior numerically, we varied the offset $a$ of the confinement
interaction, so that we can calculate the two-photon width of the pion as a
function of the resulting bound-state mass. In the mass region between 70 MeV
and 700 MeV the two-photon width can be reproduced almost quantitatively by the
function
\begin{equation}
\Gamma (M\rightarrow\gamma\gamma) = 2.95\cdot 10^{-6}eV\;
\left(\frac{M}{\mbox{MeV}} \right)^{2.95}
\end{equation}
The $M$-dependence that is expected from a simple Lagrangian thus naturally
emerges from the Salpeter formalism, in contrast to nonrelativistic
calculations.

In Tables~\ref{tabggiv} and~\ref{tabggis} we compared our results for the fit
with the instanton induced interaction by Klempt et al.~\cite{KMMP95} and the
two fits with OGE interaction to the available experimental data. Flavor mixing
is included only for the mesons with spin zero in~\cite{KMMP95}. In the other
cases the nonstrange isoscalar states in Table~\ref{tabggis} can be identified
from their isovector partners in Table~\ref{tabggiv}. 

There exist only few data for mesons, which are well established $q\bar q$
resonances. Amongst them are the tensor mesons $a_2$, $f_2$ and $f_2'$ where
our model predictions are in almost quantitative agreement with the
experimental data, which we consider a main success of this work.  In addition,
our calculation shows that the nonstrange radial excitations of the $a_2$ and
the $f_2$, namely the f-wave at around 1800 MeV and the p-wave at around 1900
MeV, one of them probably corresponding to the $f_2(1810)$, both have a
two-photon width of approximately 1 keV and therefore could be found also
experimentally. This in turn would help to clarify the nature of the various
tensor mesons and probably allow to identify glueball candidates in this
sector.

As for the pseudoscalar ground states $\pi$ and $\eta$, it has been shown in
\cite{Mue94,Mue94b}, that a quantitative description of these mesons requires
a very light quark mass of 170 MeV, with an ansatz which accounts for the
$\pi$-$\eta$ splitting and the $\eta$-$\eta'$ mixing as given by instanton effects.
Nevertheless the quark masses that have been used here give a better mass
spectrum with reasonable two-photon widths for the $\pi$, and also for the
$\eta$ and $\eta'$ meson in the model of Klempt et al.\cite{KMMP95}. The OGE
interaction of course does not give the $\eta$-$\eta'$ mixing so that these
values are far off mainly due to the wrong bound-state mass.

Our results may also help to clarify the nature of the higher lying $\eta$
resonances. The situation again is different for a
conventional OGE or an instanton induced interaction. The former predicts an
$n\bar n$ state at around 1350 MeV with a two-photon width of 650$\pm$300 eV
and an $s \bar s$ state at around 1700 MeV with a very small width. The
instanton induced interaction on the other hand predicts an almost SU(3) flavor
octet at 1550 MeV with a small width of 40 eV and a singlet at 1800 MeV with a
large width of 500 eV. An experimental investigation of the $\eta$ resonances,
especially the $\eta$(1295), could probably reveal the relevance of instanton
effects for these mesons.

The interpretation of the scalar meson spectra is nontrivial due to the
appearance of many states which do not fit into conventional $q\bar q$ nonets
-- see for instance \cite{KMMP95} for a discussion in the framework of a
relativistic quark model with instanton induced $q\bar q$-forces.  However,
both the OGE and the instanton induced interaction predict a $f_0$ $q\bar q$
ground state with a two-photon width of around 1500 eV, which is a factor of
three large than the experimental value of 560$\pm$110 eV for the $f_0(980)$,
which strengthens its interpretation as a $K\bar K$ molecule, see also
\cite{Ba85}.  Our
large two-photon would fit much better to the broad $f_0(1300)$
resonance, being probably the $q\bar q$ ground state.

In the case of the $a_0(980)$ we find a discrepancy of a factor of two between
our OGE calculations and the experimental value -- assuming that its total
width is almost equal to its partial width into $\pi\eta$. The experimental
evidence for a $K\bar K$ molecule is thus less stringent for the $a_0$ than for
the $f_0$.

As for the comparison with nonrelativistic results, we find that the ratio of
15/4\,=\,3.75 between the $0^{++}$ and the $2^{++}$ width is dramatically
reduced for light mesons -- after phase space correction -- to around 1.3 $\pm$
0.2, a phenomenon that has already been predicted by \cite{Li91}.

One of the unsolved problems in the physics of mesonic two-photon decays is the
large width of the $\pi_2(1670)$.  The nonrelativistic ansatz of Anderson et
al.~\cite{AAC91} yields very small widths of a few eV.  All the parameter sets
used in our Salpeter model predict a width around 100 eV, which is in agreement
with the calculation of \cite{AB92}, whereas the particle data booklet gives
1350$\pm$260 eV.  However, this value comes from an incoherent analysis of
different $\pi_2$ decay modes, whereas it has been convicingly argued in
\cite{cello89} to use a constructive interference between the $f_2\pi$ and
$\rho\pi$ decay mode, which yields 800$\pm$300$\pm$120 eV. Recent results from
\cite{ARG94} also give a smaller value of $470^{+140} _{-180}\, $eV, which is
almost compatible with the quark model calculations. A possible explanation of
the discrepancy between theory and experiment would be an interference between
the $\pi_2$ with the second $a_2$ state, which up to now has not been found,
but within quark models should have a mass of around 1800 MeV. As in our
calculation it has a width of 300 eV, it could have influenced the two
experiments cited by the Particle Data Group~\cite{PDG94}, as none of them has
performed an analysis of angular distributions.

Finally we would like to emphasize those mesons which can most probably be
seen in $\gamma\gamma$ experiments because of their large two-photon width,
namely the $a_2$ and the nonstrange $f_2$ including their radial excitations,
the $a_0$ and $f_0$ ground states and the radial excitation of the nonstrange
$\eta$. However, in has been found that also angular excited mesons such as
the nonstrange $f_3$ and $f_4$ or the $\eta_2$ are also promising candidates.

\section{Summary and Conclusion} \label{V}
We have presented a relativistic quark model calculation including a complete
numerical study of the annihilation on quark-antiquark states into two
photons. Several conceptional aspects have been improved with respect to
previous theoretical work: The meson mass spectra are calculated in the
framework of the Salpeter equation, which includes negative energy components
of the meson wave function. These allow for a normalization condition given by
Salpeter, which is adequate for states with large binding energy, as is the
case for light quarkonia.

The meson to two-photon transition matrix element is calculated in the
Mandelstam formalism. To this aim we use four-dimensional Bethe-Salpeter
amplitudes which have been reconstructed from the Salpeter amplitudes that are
the solutions of the bound-state mass eigenvalue equation. In contrast to
previous work we thus do not put the quarks on mass-shell in the transition
matrix element, but evaluate the four-dimensional Feynman diagram including the
dependence on the relative energy. Together with the Salpeter normalization
this leads for instance in the case of pseudoscalars to a reasonable behavior
$\Gamma^{}_{M\rightarrow \gamma\gamma} \sim M^3$ as expected from a simple
Lagrangian. Therefore our model does not rely on the additional ``Mock-meson''
factor which has been necessary in previous work and contains an additional
theoretical uncertainty in the scale of the two-photon widths.

In the numerical study that has been presented in the second part of this paper
we have compared the parameter sets of non- and semi-relativistic quark model
and of Salpeter models for the meson spectra. The results for most of the
two-photon widths are found to be stable against parameter changes.  From the
various models we have extracted theoretical predictions including error
estimates for the widths of singlet and triplet quark-antiquark states up to
spin six.

As our results agree almost quantitatively with many well established $q\bar q$
resonances such as the $\eta_c,\;\chi_{c0},\;\chi_{c2}$ or the light tensor
nonet $a_2,\;f_2$ and $f_2'$, it is hoped that this complete numerical study
can be used as a guideline for present and future experiments. In particular,
we suggest that the most successful candidates for an experimental
investigation in the heavy quark sector are the $\eta_c$ and $\chi_0$ including
their radial excitations. In the light quark sector, the nonstrange $f_2$ and
its radial excitations couple most strongly to two photons. Both the second p-
and the first f-wave state have a width of around 1 keV and therefore may be
measured experimentally.  This in turn would clarify the $f_2$ meson mass
spectrum and could probably allow the identification of tensor glueballs, one
of the most interesting QCD phenomena.


\begin{table}
  \caption{Parameters and interaction type of the different models}
  \label{tabparams}
  \centering
   \begin{tabular}{lcccccc}
                   & Quark masses                & Confinement &  & & Residual
   & \\
   Model           & $m_n\;\;\;m_s\;\;\;m_c\;\;\;\;\,m_b$  & $\!$Spin
   structure$\!$ & $a$ &  $b$ & Interaction & $\alpha_{sat}$ \\
\hline
   Godfrey \& Isgur~\protect{\cite{GI}} & $\;\;-\;\;\;\; -\;\; 1628 \;\;\; 4977$ &
   Scalar & -253 & 914 & OGE-Coul-Run & 0.60 \\ 
   Bhaduri et al.~\protect{\cite{BC}} & $\;\;-\;\;\;\; -\;\; 1870 \;\;\; 5259$
   & Scalar &-913 &941 & OGE-Coul-Fix &0.39  \\
   Resag \& M\"unz V~\protect{\cite{RM95}}$\!\!\!$  & $\;\;-\;\;\;\; -\;\; 1631
  
   \;\;\; 5005$ & Vector & -640 & 1291 & OGE-Coul-Run & 0.365 \\
   Klempt et al.~\protect{\cite{KMMP95}} & $\!\!\!\!306\;\;503\;\;\;\;
   -\;\;\;\; -$ & Scal.+Vect. 
   & -1751 & 2076 & 't Hooft & - \\ 
   OGE-SRM & $\!\!\!\!220\;\;468\;\;\;\; -\;\;\;\; -$  & $\!$Scal.+Vect.$\!$ &
   -1273 &  1856 & OGE-Feyn-Run & 0.224 \\
   OGE-NRM & $\!\!\!\!330\;\;567\;\;\;\; -\;\;\;\; -$  & $\!$Scal.+Vect.$\!$ &
   -1418 &  1956 & OGE-Coul-Run & 0.303 \\
\end{tabular}
\end{table}

\begin{table}
  \caption{Calculated $\gamma\gamma$-widths of charmonium states in eV 
    for several models and the theoretical estimate compared to experimental
    data (meson masses in GeV are given in parenthesis).}
  \label{tabggcc}
  \centering
   \begin{tabular}{lcccccccc}
       & $J^{PC}$ & Godfrey~\protect{\cite{GI}} & Bhaduri~\protect{\cite{BC}} &
       Resag V~\protect{\cite{RM95}} & Theor.\ Estimate$\!\!\!$ & Exp.  \\ 
     \hline
     $\eta_c$(2.98)  & $0^{-+}$ & 3690 (2.97)  & 2960 (2.88)   &
     3820 (2.98)  &  3500 $\pm$ 400 & 
                         7000$^{+ 2000}_{- 1700}$  ~\protect{\cite{PDG94}} \\
          & & & & & &   4300$\pm$1900~\protect{\cite{Cleo95}} \\
     $\eta'_c$(3.59) & $0^{-+}$ & 1400 (3.62)  & 1000 (3.57) 
     & 1730 (3.67) & 1380$\pm$300  &   \\
     $\eta''_{c}$ & $0^{-+}$    & 930 (4.02) & 657 (3.99) & 1220 
     (4.16)  & 940$\pm$230  &  \\
     $\eta'''_{c}$ & $0^{-+}$   & 720 (4320)  & 502 (4.32)  & 979 (4.58)  &
     730$\pm$200  &  \\ 
     $\eta''''_{c}$ & $0^{-+}$  & 610 (4580)  & 421 (4.60)  & 831 (4.94) &
     620$\pm$170  &  \\ 
     $\eta_{c2}$  & $2^{-+}$    & 9.1 (3.77) & 4.1 (3.79)  &13.6 (3.84)&
     9$\pm$4  &  \\ 
     $\eta'_{c2}$ & $2^{-+}$    & 12.1 (4.11)  & 7.2 (4.14) & 20.2 (4.28) &
     13$\pm$6 & \\ 
     $\eta_{c4}$  & $4^{-+}$    & 0.108 (4.16) & 0.028 (4.19)    & 0.304
     (4.40)  & 0.15$\pm$0.12  &  \\
     \hline
     $\chi_{c0}$(3.41)  & $0^{++}$    & 1290 (3.45) & 1270 (3.49) & 1620 (3.40)&
     1390$\pm$160 &   4000$\pm$2800  ~\protect{\cite{PDG94}}  \\ 
          & & & & & &   1700$\pm$800~\protect{\cite{Cleo95}} \\
     $\chi'_{c0}$  & $0^{++}$   & 950 (3.88) & 1140 (3.90) & 1250 (3.94) &
     1110$\pm$130  &  \\
     $\chi''_{c0}$ & $0^{++}$ & 741 (4.20) & 969 (4.23) & 1020 (4.38) &
     910$\pm$130  &  \\ 
     $\chi_{c2}$(3.55)   & $2^{++}$   & 459 (3.53) & 259 (3.54) & 601 (3.56)& 440
     $\pm$ 140 &321$\pm$95 ~\protect{\cite{PDG94}}  \\ 
         & & & & & &   700$\pm$300~\protect{\cite{Cleo95}} \\
     $\chi'_{c2}$  & $2^{++}$   & 449 (3.93) & 317 (3.95) & 684 (4.05)&
     480$\pm$160  &   \\ 
     $\chi''_{c2}$ & $2^{++}$& 16.3 (3.99) & 3.7 (4.02)  & 23.4 (4.09) &
     14$\pm$8  &  \\ 
     $\chi_{c3}$ & $3^{++}$  & 1.53 (4.00)  & 0.44 (4.02)  &3.07 (4.13)
     & 1.7$\pm$1.1  &    \\
     $\chi_{c4}$ & $4^{++}$  & 1.09 (3.99)  & 0.31 (4.01) &2.12 (4.19)   &
     1.2$\pm$0.8  &   \\ 
\end{tabular}
\end{table}

\begin{table}
  \caption{Calculated \protect{$\gamma\gamma$}-widths of bottomonium states in
    eV for several models.}
  \label{tabggbb}
  \centering
   \begin{tabular}{lcccccc}
       & $J^{PC}$ & Godfrey~\protect{\cite{GI}} & Bhaduri~\protect{\cite{BC}} &
       Resag V~\protect{\cite{RM95}}  & Theor. Estimate   \\ 
     \hline
     $\eta_b$ & $0^{-+}$     & 214 (9.47)  & 266 (9.37)   & 192 (9.44)&
     220$\pm$40  \\ 
     $\eta'_b$ & $0^{-+}$    & 121 (10.01)   & 95.0 (10.01)   & 116 (10.01) &
     110$\pm$20  \\ 
     $\eta''_{b}$ & $0^{-+}$ & 90.6 (10.35)  & 67.9 (10.36)  & 93.5 (10.39) &
     84$\pm$12 \\ 
     $\eta'''_{b}$ & $0^{-+}$ & 75.5 (10.62) & 56.3 (10.63)  & 81.8 (10.72) &
     71$\pm$11 \\ 
     $\eta_{b2}$ & $2^{-+}$  & 41.6 meV (10.13)  & 28.3 meV (10.19) &  51.3 meV
     (10.13) & 40$\pm$10 \\ 
     $\eta'_{b2}$ & $2^{-+}$ & 69.8 meV (10.43)  & 52.3 (10.47)  & 96.2 meV
     (10.47) & 73$\pm$18 \\ 
     $\eta_{b4}$  & $4^{-+}$ & 41.0 $\mu$eV (10.51)& 15.9 $\mu$eV (10.54)  &
     71.9 $\mu$eV (10.56) & 43$\pm$23 \\ 
     \hline
     $\chi_{b0}\,$ (9.86) & $0^{++}$  & 20.8 (9.89) & 27.3 (9.91) & 24.1 (9.83) &
     24$\pm$3  \\ 
     $\chi'_{b0}$(10.23)& $0^{++}$ & 22.7 (10.25) & 26.9 (10.27) & 27.3 (10.25)&
     26$\pm$2\\ 
     $\chi_{b2}\,$ (9.91) & $2^{++}$ & 5.14 (9.92)& 5.26 (9.95) & 6.45 (9.87) &
     5.6$\pm$0.6 \\ 
     $\chi'_{b2}$(10.27)& $2^{++}$ & 6.21 (10.27) & 6.11 (10.30) & 8.1 (10.28)&
     6.8$\pm$1.0\\ 
     $\chi_{b3}$& $3^{++}$   & 1.57 meV (10.35) & 0.72 meV (10.39)   & 2.41
     meV (10.36) & 1.6$\pm$0.7  \\ 
     $\chi_{b4}$& $4^{++}$   & 1.26 meV (10.35) & 0.58 meV (10.39)   & 1.94
     meV (10.37) &  1.3$\pm$0.6 \\
\end{tabular}
\end{table}

\begin{table}
  \caption{Comparison of the scalar and tensor charmonium and bottomonium widths
    with the results of Bergst\"om et al.~\protect{\cite{Ber83}} }
  \label{tabberg}
  \centering
   \begin{tabular}{cccccc}
       & $J^{PC}$ & Salpeter-model & Bergst\"om~\protect{\cite{Ber83}} &
       Bergst\"om non-rel.~\protect{\cite{Ber83}}  &
       Exp. \\  
     \hline
     $\chi_{c0}$  & $0^{++}$    & 1270 & 1360 & 2960 &
     1700$\pm$800~\protect{\cite{Cleo95}} \\ 
     $\chi_{c2}$   & $2^{++}$   & 259  & 498 & 789 &
     321$\pm$95~\protect{\cite{PDG94}} \\ 
     \hline
     $\chi_{b0}$ & $0^{++}$  & 27.3 & 27.6 & 44.3 &    \\
     $\chi_{b2}$ & $2^{++}$  & 5.26  & 9.35 & 11.8 & \\
\end{tabular}
\end{table}

\begin{table}
  \caption{Calculated \protect{$\gamma\gamma$}-widths of isovector states in eV 
    for a model with instanton induced~\protect{\cite{KMMP95}} and with OGE
    interaction compared to experimental
    data (meson masses in MeV are given in
  parentheses).}
  \label{tabggiv}
  \centering
   \begin{tabular}{lccccc}
       & $J^{PC}$ & Klempt~\protect{\cite{KMMP95}} & OGE-NRM
        & OGE-SRM & Exp.   \\ 
     \hline
       $\pi\,$ (138)  &$0^{-+}$ & 4.23 (140) & 3.81 (140) & 5.07 (140)  &
       7.84$\pm$0.56 \cite{PDG94}\\ 
       $\pi'$(1300) &$0^{-+}$ & 151 (1360)& 127 (1380)  & 355 (1330)  & \\
       $\pi''$ &$0^{-+}$ & 2.00 (2010)& 4.09 (2020) & 7.47 (1920)  & \\
       $\pi_2$(1670)&$2^{-+}$ & 94.2 (1630)& 73.2 (1650) & 129 (1560)  &
       1350$\pm$260 \cite{PDG94}\\ 
               &    &   &   &  & 470$^{+ 140}_{- 180}$~\protect{\cite{ARG94}} \\
       $\pi'_2$(2100)&$2^{-+}$ & 12.0 (2160)& 4.93 (2150) & 2.48 (2020)  &  \\
       $\pi_4$&$4^{-+}$ & 28.3 (2220)& 23.6 (2260) & 51.2 (2120)  & \\
       $\pi_6$&$6^{-+}$ & 10.4 (2680)& 8.4 (2720) & 20.8 (2550)  & \\
     \hline
       $a_0\;$ (980)  &$0^{++}$ & 1390 (1320)& 640 (1080) & 486 (1010)   &
       ${}^> _\sim$240$^{+80}_{-70}$ \cite{PDG94}\\
       $a'_0$ &$0^{++}$ & 386 (1930)& 54.0 (1780) & 28.5 (1680)  &\\
       $a''_0$ &$0^{++}$ & 291 (2420) & 30.3 (2290) & 11.8 (2140)  &\\
       $a_2$(1320)  &$2^{++}$ & 734 (1310)& 766 (1330) & 900 (1280)  &
       1040$\pm$90 \cite{PDG94} \\
       $a'_2$ &$2^{++}$ & 374 (1880)& 326 (1870) & 376 (1740)  & \\
       $a''_2$ &$2^{++}$& 293 (1930)& 320 (1930) & 353 (1820) & \\
       $a_3$  &$3^{++}$ & 60.9 (1950)& 56.2 (1970) & 93.5 (1840)  & \\
       $a_4$(2040)  &$4^{++}$ & 50.4 (2010)& 43.8 (2030)& 87.9 (1910)  & \\
       $a_5$  &$5^{++}$ & 15.8 (2460)& 13.5 (2490) & 27.9 (2320)  & \\
       $a_6$(2450)  &$6^{++}$ & 14.0 (2520)& 11.2 (2530)& 28.5 (2380)  & \\
\end{tabular}
\end{table}

\begin{table}
  \caption{Calculated \protect{$\gamma\gamma$}-widths of isoscalar light
    quarkonia in eV.}
  \label{tabggis}
  \centering
   \begin{tabular}{lccccc}
       & $J^{PC}$  & Klempt~\protect{\cite{KMMP95}} & OGE-NRM
        & OGE-SRM & Exp.~\protect{\cite{PDG94}}   \\ 
     \hline
       $\eta$ (547) & $0^{-+}$  & 208 (530) & 10.6 (142) & 14.1 (138)  &
       460$\pm$40     \\ 
       $\eta'$(958) &$0^{-+}$ & 2330 (980) & 39.5 (665) & 45.8 (641)  &
       4260$\pm$190     \\ 
       $\eta''$(1295) &$0^{-+}$& 31.8 (1530)& 350 (1390) & 986 (1330)  &  \\
      $\eta'''$ &$0^{-+}$& 499  (1810)& 0.75 (1700) & 0.27 (1660)  & \\
     $\eta''''$ &$0^{-+}$& 510  (2180)& 11.4 (2020) & 20.8 (1920)  & \\
    $\eta'''''$ &$0^{-+}$& 544  (2380)& 5.7 (2350) & 3.9 (2280)  & \\
       $\eta_2$ &$2^{-+}$& 262 (1630)& 203 (1650) & 358 (1560)  & \\
       $\eta'_2$ &$2^{-+}$& 9.89 (1860)& 7.2 (1940) & 10.7 (1870)  & \\
       $\eta''_2$&$2^{-+}$ & 33.4 (2160)& 13.7 (2150) & 6.9 (2020)  &  \\
       $\eta'''_2$&$2^{-+}$ &4.22 (2420) & 2.7 (2490) & 2.9 (2400)  &  \\
       $\eta_4$ &$4^{-+}$& 78.7 (2220)& 65.7 (2260)& 142 (2120)  & \\
       $\eta'_4$ &$4^{-+}$& 1.77 (2480)& 1.25 (2580) & 22.0 (2470)  & \\
       $\eta_6$&$6^{-+}$& 28.9 (2680)& 23.4 (2720) & 57.7 (2550)  & \\
       $\eta'_6$&$6^{-+}$& 0.46 (2960)&10.3 (3050)  & 17.2 (2840)  & \\
     \hline
       $f_0\,$ (980)   &$0^{++}$ & 1750 (980) & 1780 (1080)  & 1350 (1010)  &
       560$\pm$110   \\ 
       $f_0$(1300)  & $0^{++}$   &   &  &   & 5400$\pm$2300  \\
       $f'_0$(1590)  &$0^{++}$ & 161 (1470)& 192 (1400) & 199 (1360)  & \\
       $f''_0$ &$0^{++}$ & 15.3 (1780)& 150 (1780) & 79 (1680)  & \\
       $f'''_0$ &$0^{++}$& 29.2 (2110)& 50.6 (2110) & 45.5 (2050)  & \\
       $f''''_0$ &$0^{++}$& 42 (2310)& 84.3 (2290) & 32.7 (2140) & \\
       $f_2$(1270)&$2^{++}$ & 2040 (1310)& 2130 (1330) & 2500 (1280)  &
       2440$^{+ 320}_{- 290}$  \\ 
       $f'_2$(1525)  &$2^{++}$ & 121 (1525)& 127 (1590) & 146 (1540)  &
       105$\pm$17 \\ 
       $f''_2$(1810) &$2^{++}$ & 1038 (1880)& 906 (1870) & 1040 (1740)  & \\
       $f'''_2$ &$2^{++}$& 815 (1930)& 888 (1930) & 982 (1820)  & \\
       $f''''_2$(2010) &$2^{++}$& 75.2 (2150)& 83.8 (2220)& 52.9 (2130) & \\
       $f'''''_2$(2300) &$2^{++}$& 42.7 (2160) & 33.7 (2230) & 75.9 (2160) & \\
       $f_3$   &$3^{++}$ & 169 (1950)& 156 (1970) & 260 (1840)  & \\
       $f'_3$  &$3^{++}$ & 5.38 (2190)& 4.56 (2290) & 6.60 (2190)  & \\
       $f_4$(2050)   &$4^{++}$ & 140 (2010)& 122 (2030) & 244 (1910)  & \\
       $f'_4$(2220)  &$4^{++}$ & 3.81 (2230)& 3.07 (2330) & 121 (2240)  & \\
       $f_5$   &$5^{++}$ & 44.0 (2460)& 37.4 (2490) & 77.5 (2320)  & \\
       $f'_5$  &$5^{++}$ & 0.94 (2730)& 0.67 (2830) & 1.15 (2700)  & \\
       $f_6$(2510)   &$6^{++}$ & 38.9 (2520)& 31.0 (2530) & 79.1 (2380)  & \\
       $f'_6$  &$6^{++}$ & 0.654 (2770)& 0.435 (2860) & 0.827 (2740)  & \\
\end{tabular}
\end{table}

\end{document}